\documentclass[10pt]{article}  
\usepackage{color,amssymb,amsthm,amsmath,amsxtra,amsfonts,breqn,
fullpage,graphicx,hyperref,float,appendix,mathtools,eqparbox, 
lscape,pdfpages,comment,multicol,bm}

\usepackage[cal=boondox]{mathalfa}

\usepackage{txfonts}

\usepackage[latin1]{inputenc}
\usepackage{tikz}
\usetikzlibrary{shapes,arrows}

\usepackage{fancyhdr}
\usepackage{xcolor}
\definecolor{dark-red}{rgb}{0.4,0.15,0.15}
\definecolor{dark-blue}{rgb}{0,0,0.45}
\definecolor{dark-green}{rgb}{0.008,0.392,0.251}
\hypersetup{colorlinks, linkcolor={dark-blue},citecolor={blue}, urlcolor={dark-blue}}

\numberwithin{equation}{section}
\usepackage{pdflscape}

\definecolor{gold}{rgb}{0.85,0.66,0.0}
\def\todo#1{\textcolor{red}{\textbf{**** TODO -- #1 ****}}}
\def\writeme#1{\textcolor{dark-green}{\textbf{WRITE ME!! \qquad #1}}}

\theoremstyle{definition}

\usepackage{etoolbox}
\makeatletter
\providecommand{\institute}[1]{
  \apptocmd{\@author}{\end{tabular}
    \par
    \begin{tabular}[t]{c}
    #1}{}{}
}
\makeatother

\usepackage[ddmmyyyy,hhmmss]{datetime}

\begin{document}


\title{Embedded Blockchains: A Synthesis of Blockchains, Spread Spectrum Watermarking, %
Perceptual Hashing \& Digital Signatures}
\author{Sam Blake\\ \href{mailto:sam.blake@unimelb.edu.au}{sam.blake@unimelb.edu.au}}
\institute{\textit{The University of Melbourne}}
\maketitle

\begin{abstract}
In this paper, we introduce a scheme for detecting manipulated audio and video. The scheme is a synthesis of 
blockchains, encrypted spread spectrum watermarks, perceptual hashing and digital signatures, which we call an 
Embedded Blockchain. Within this scheme, we use the blockchain for its data structure of a cryptographically 
linked list, cryptographic hashing for absolute comparisons, perceptual hashing for flexible comparisons, 
digital signatures for proof of ownership, and encrypted spread spectrum watermarking to embed the blockchain 
into the background noise of the media. So each media recording has its 
own unique blockchain, with each block holding information describing the media segment. The 
problem of verifying the integrity of the media is recast to traversing the blockchain, 
block--by--block, and segment--by--segment of the media. If any chain is broken, the 
difference in the computed and extracted perceptual hash is used to estimate the level of manipulation. 
\end{abstract}

\textbf{Index Terms} -- blockchains, digital watermarking, perceptual hashing, digital 
signatures, tamper detection, deepfakes.

\begin{multicols}{2}

%
%

\section{Introduction}

The creation of artificially--generated video and audio has become more prevalent in recent years with 
advances in deep learning technologies including generative adversarial networks (GANs) and are now at a 
point where they are staggeringly realistic \cite{zhang2019}\cite{nguyen2019}. These are commonly known 
as \textit{deepfakes}. \\

Deep learning methods for generating deepfakes are improving rapidly. In the future it may be 
impossible to distinguish deepfakes from real media\footnote{Hao Li, associate professor at the University of Southern California and CEO of Pinscreen was said ``At some point it's likely that it's not going to be possible to detect [AI fakes] at all. So a different type of approach is going to need to be put in place to resolve this.'' \cite{vincent2019}.}. The majority of the methods proposed to 
detect deepfakes involve using deep learning \cite{nguyen2019}\cite{li2020}. \\

There is an ever--increasing commercial desire to track the provenance and integrity of audio, 
photos and video. In 2013, IEEE Information Forensics and Security Technical Committee (IFS--TC) hosted 
the first Image Forensics Detection Challenge \cite{ifstc2013}. Ongoing notable projects aiming to address 
deepfakes or manipulated video and audio include 
\href{https://ai.facebook.com/datasets/dfdc/}{The Deepfake Detection Challenge} by Facebook and 
Microsoft, The News Provenance Project by \textit{The New York Times} and the DARPA MediFor 
(Media Forensics) program \cite{dolhansky2019}\cite{medifor}\cite{nyt}. \\

Contemporary solutions for detecting deepfakes or manipulated media generally operate  
after the suspected manipulations have taken place. We propose a two--step process; firstly, the 
media is marked in real--time as its captured with a series of unique identifiers. Subsequently, 
if the provenance of the media is questioned, the second step of verification of occurs. \\

In order to verify the integrity of the media using only the media and its embedded data, the unique 
identifiers should possess the following properties:
\begin{enumerate}
\item Hidden from human perception.
\item Provide proof of ownership.
\item Describe the content.
\item Impose a temporal ordering on the content. 
\end{enumerate}

These identifiers could be hidden (or partially hidden) from human perception using a spread 
spectrum watermark \cite{cox1997}\cite{cox2007}\cite{shih2017}\cite{tirkel1993}. \\

Proof of ownership can be verified by extraction of the encrypted watermark data, or by 
verification of a digital signature once the watermark data is extracted and decrypted. \\

Cryptographic hashing can provide a bit--level description of the content, and perceptual 
hashing could provide a flexible description of the content. \\

Blockchains are used to link segments of the media together and impose a temporal ordering via hashing. The 
blockchain also prevents the removal, insertion or rearranging of any segments (blocks) of the media. \\ 

We call the synthesis of blockchains, encrypted spread spectrum watermarking and perceptual hashing an 
\textit{Embedded Blockchain}. \\

Our solution does not require a trusted authority to verify the media stream against an original, 
as all the necessary data describing the media is contained within the Embedded Blockchain which 
is hidden in the media. \\

An obvious application of Embedded Blockchains is smartphones, where hidden datasets 
could be embedded on the fly into the video and audio recordings. The Embedded Blockchain 
could provide a unique link to the smartphone via the digital signature.


%
%

\section{Blockchains}

Blockchains were invented in 2008 by a person or group of people, under the name Satoshi 
Nakamoto \cite{nakamoto2008}. The blockchain serves as a public transaction ledger of the 
cryptocurrency, Bitcoin. Nakamoto outlined a data structure for chaining blocks of data 
together by their cryptographic hash. Previously, Haber and Stornetta developed a robust 
method for time-stamping digital documents by hashing the document with that of the previous 
document, thus making a cryptographically linked list \cite{haber1991}. \\

The central idea is any changes to the data in any one block will, with overwhelming 
probability, change the hash of the block and all subsequent blocks; thus imposing an 
immutable temporal ordering of the data in the blocks.

\begin{figure}[H]
\centering
\includegraphics[width=0.45\textwidth]{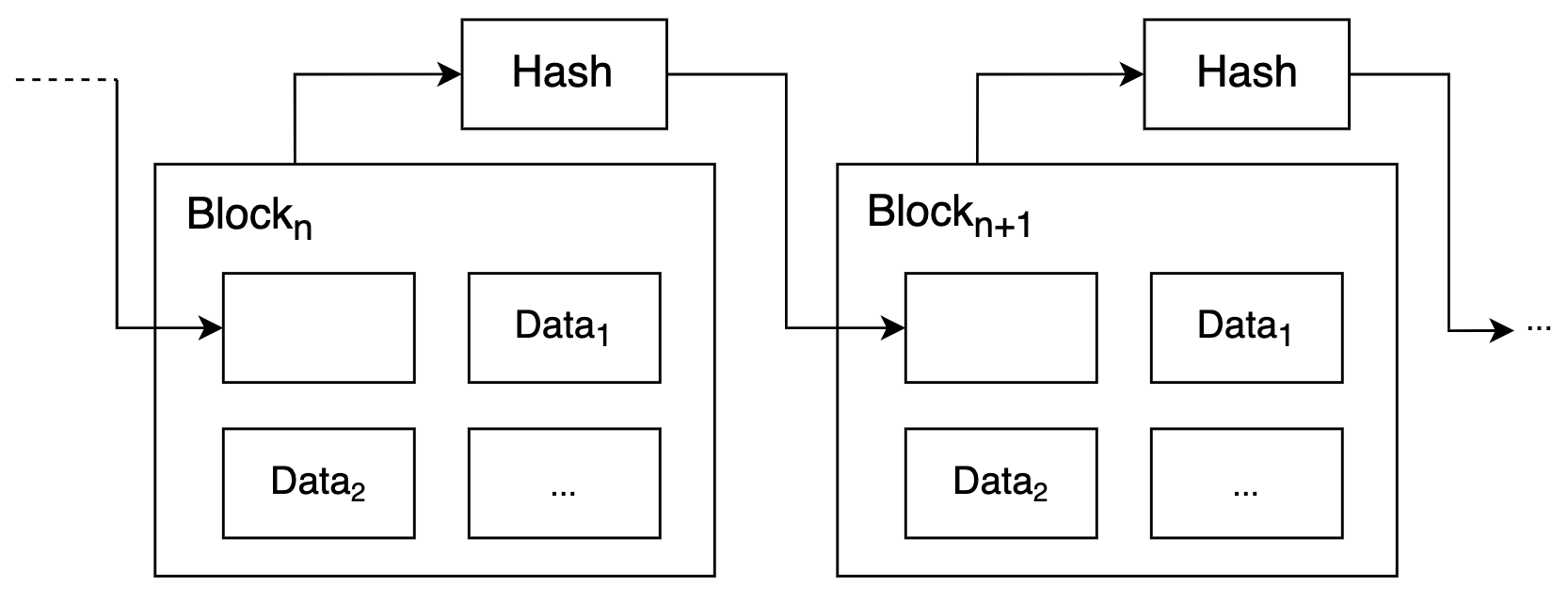}
\caption{A conceptual diagram of the blockchain.}
\end{figure}

Our use of the blockchain is primarily for its data structure. For our purposes, we can think of 
the blockchain as a cryptographically linked list. However, we generalise the links used to 
chain blocks together by including both a cryptographic and perceptual hash.

\section{Spread Spectrum Watermarking}

Spread spectrum (Digital) watermarking (or watermarking) is a steganographic technique which embeds an 
imperceptible message in a noise-tolerant signal \cite{cox1997}\cite{cox2007}\cite{shih2017}\cite{tirkel1993}.  
Spread spectrum watermarking was invented by Andrew Tirkel and Charles Osborne in 1992 \cite{tirkel1993}. Commonly 
watermarked noise--tolerant signals include audio, imagery and video. More abstract datasets are also 
amenable to watermarking \cite{panah2016}. \\	

The hidden message is embedded into the signal using families of sequences, or multi-dimensional arrays 
which possess good periodic autocorrelation and pairwise cross--correlation \cite{blake2013}\cite{blake2020}\cite{tirkel2015}. 
The message is encoded as cyclic shifts of the sequences or multi--dimensional arrays. \\

A secure watermark can be created by encrypting the families watermarking arrays prior to 
embedding in the signal. In recent years a number of image--based encryption methods have been 
proposed \cite{chen2009}\cite{tang2016}\cite{wang2017}\cite{zheng2015}. \\

The process of embedding a watermark in a media stream is summarised in the following diagram. 

\begin{figure}[H]
\centering
\includegraphics[width=0.4\textwidth]{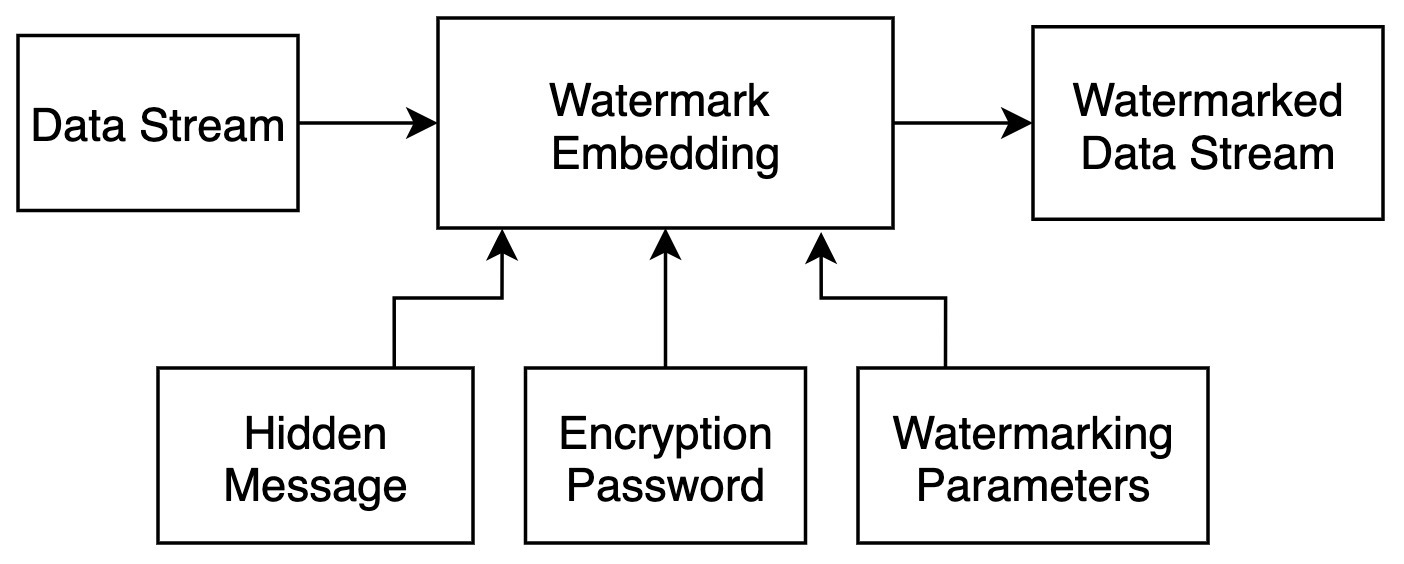}
\caption{A conceptual diagram of the watermark embedding process.}
\end{figure}

In the diagram above, the watermarking parameters may include the embedding strength, the 
embedding domain, the family of arrays, the array size and dimensionality of the arrays. \\

The number of arrays required to embed the hidden message depends on the message length, 
the array size and array dimensionality (as increasing the dimensionality increases the 
watermark payload capacity \cite{blake2020}). Estimating the optimal embedding strength requires 
balancing perceptibility of the watermark with robust extraction of the hidden message. 
Furthermore, the robust extraction of the watermark is dependent on the number of arrays 
embedded and consequently depends on the size of the hidden message.\\

The extraction of a watermark is summarised in the following diagram. 

\begin{figure}[H]
\centering
\includegraphics[width=0.4\textwidth]{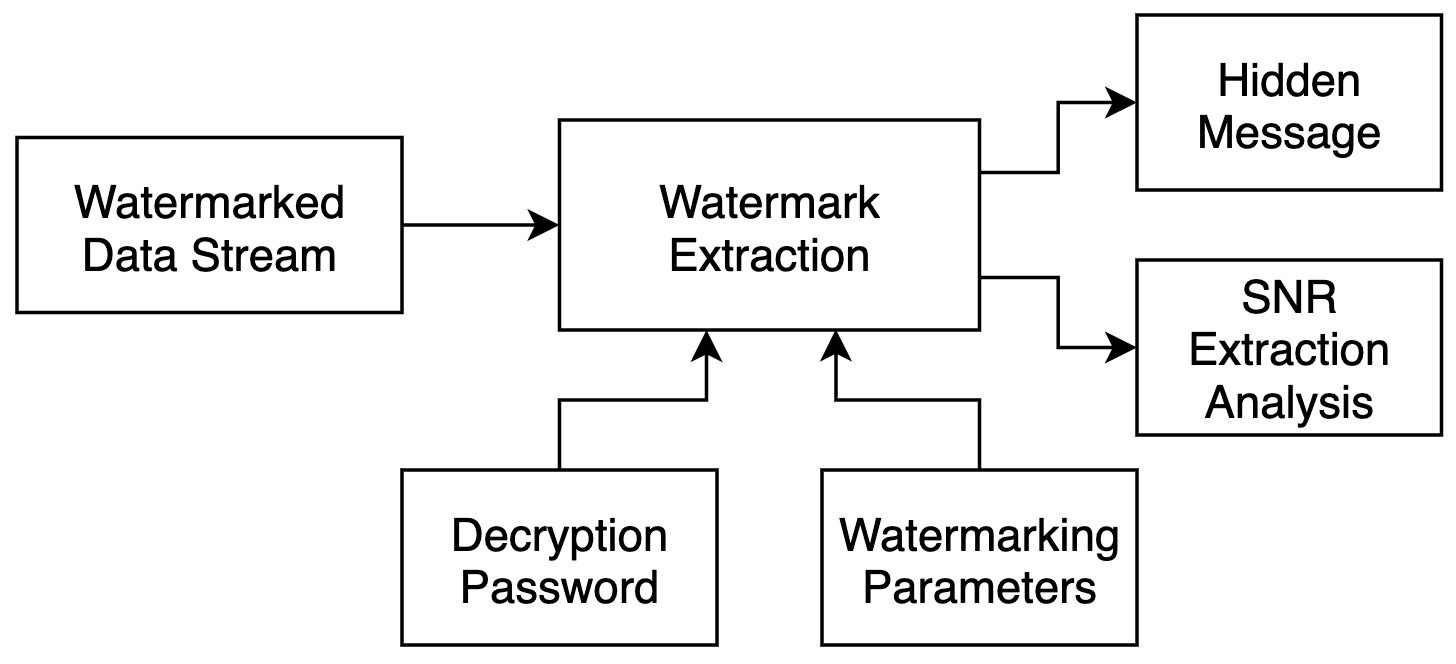}
\caption{A conceptual diagram of the watermark extraction process.}
\end{figure}

To determine if the watermark extraction was successful, the signal--to--noise 
ratio (SNR) of each array is compared to a predetermined noise level.

%
%

\section{Cryptographic \& Perceptual Hashing}

Cryptographic hash functions compress an arbitrary number of bits into a fixed 
number of bits. The cryptographic hash is a one--way function, which is 
computationally prohibitive to invert. They act as a unique identifier of the 
input data, such that two similar inputs will almost certainly result 
in vastly different outputs (by the avalanche effect \cite{feistel1973}). If 
two distinct inputs return the same cryptographic hash, known as a collision, then 
the hashing function is considered broken. Many hashing functions have been broken, 
such as the widely used MD5 hash \cite{buldas2006}. Robust cryptographic hashing 
functions include the SHA2 family, which were developed by the NSA \cite{lilly2001}. \\

Perceptual hashing functions are very different from cryptographic hashing functions. Unlike 
cryptographic hashes, perceptual hashes are similar if their inputs possess similar 
features. The difference (Hamming distance) between perceptual hashes is used as a 
metric of dissimilarity. Perceptual hashes apply to both audio and imagery 
\cite{grutzek2012}\cite{niu2008}\cite{venkatesan2000}.

%
%

\section{Digital Signatures}

Digital signatures are a form of asymmetric encryption which is used to bind a person 
or entity to digital data \cite{diffie1976}\cite{rsa1978}. We sign some data with a private 
key. Once signed, the data and the signature can be disseminated. Verification happens with a 
public key, the data and the signature. If the data or signature has changed the public key 
will invalidate the signature. \\

Incorporating a digital signature into the Embedded Blockchain gives us the ability to 
distribute the watermark encryption password, so the media can be independently verified 
and still keep control of the creation of our unique Embedded Blockchain via the private 
key of the digital signature.

%
%

\section{Embedded Blockchains}

We now have all the necessary components to describe the Embedded Blockchain. Our description 
is in terms of video, however, it can be applied to audio or any abstract dataset which has a 
temporal or lexical ordering and is amenable to spread spectrum watermarking. \\

As discussed earlier, we embed a series of groups of unique identifiers into the video with 
the following features:
\begin{enumerate}
\item Hidden from human perception.
\item Provide proof of ownership.
\item Describe the content.
\item Impose a temporal ordering on the content. 
\end{enumerate}

Each group of unique identifiers describes one segment of the media. We refer to the data 
in this group as a block. The blocks are chained together using cryptographic and 
perceptual hashes. Additionally, a small amount of user--defined data may be included in 
each block, for example, a timestamp or some covert information. \\



Before we can describe the embedding and extraction processes we need to decide on the segment 
size that we wish to describe in each block. For simplicity, we can think of the segment size as 
a single frame of video. This size can vary to accommodate real--time embedding or to balance 
embedding time with the fidelity of the perceptual hash for application--specific requirements. 

\subsection{Describe \& Embed}

Consider the first segment, which is different from all subsequent segments, as it does not 
contain a block. We compute the perceptual and cryptographic hashes of this segment and digitally 
sign the perceptual hash of the segment. Then we form the block by accumulating this data with the 
user-defined data and embed it into the second segment using the spread spectrum watermark. \\

Segments are chained together by embedding the data describing segment $n$ into segment 
$n+1$. This is summarised in the diagram below. \\

\begin{figure}[H]
\centering
\includegraphics[width=0.45\textwidth]{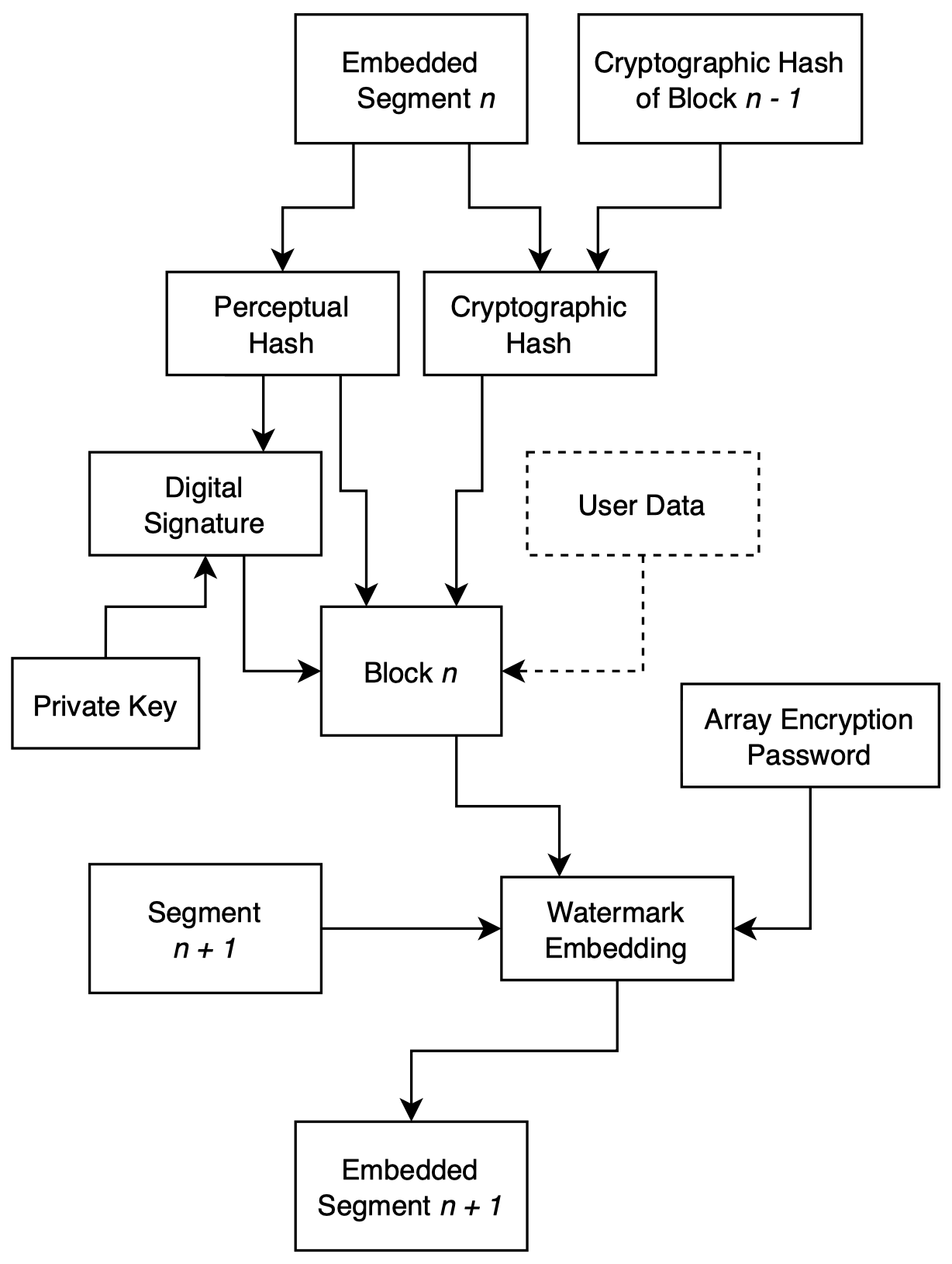}
\caption{A conceptual diagram of the creation of the Embedded Blockchain.}
\end{figure}

If we are using lossy compression to store the video, then we replace the cryptographic hash 
of the $n$--th segment with the cryptographic hash of the perceptual hash of the $n$--th 
segment. This way, modifying a single segment will still break all subsequent chains. \\


\subsection{Extract \& Verify}

Given a video which contains an Embedded Blockchain we wish to verify that the video has 
not been manipulated. As this video contains an Embedded Blockchain, we can recast the problem 
of detecting any manipulations to traversing the Embedded Blockchain, block--by--block and 
segment--by--segment, where for each segment we compute the cryptographic hashes, perceptual hashes
and digital signatures, then compare these computed values to the corresponding data 
extracted from the embedded block. If the cryptographic hash does not match, then some modification 
has occurred and we can measure the level of modification with the difference between the computed and 
extracted perceptual hashes. \\

If the block is not recoverable from the watermark, then it is immediately evident that the chain of 
blocks is broken and the segment has been modified.  \\

It is important to note that we have not relied on any external data to make this determination. For 
example, we have not compared the video to an untampered original, nor looked up the hash of the video 
in a database, nor relied upon a deep learning approach which needs millions of test cases for training. \\

The extraction and verification scheme is summarised in the following diagram. 

\begin{figure}[H]
\centering
\includegraphics[width=0.45\textwidth]{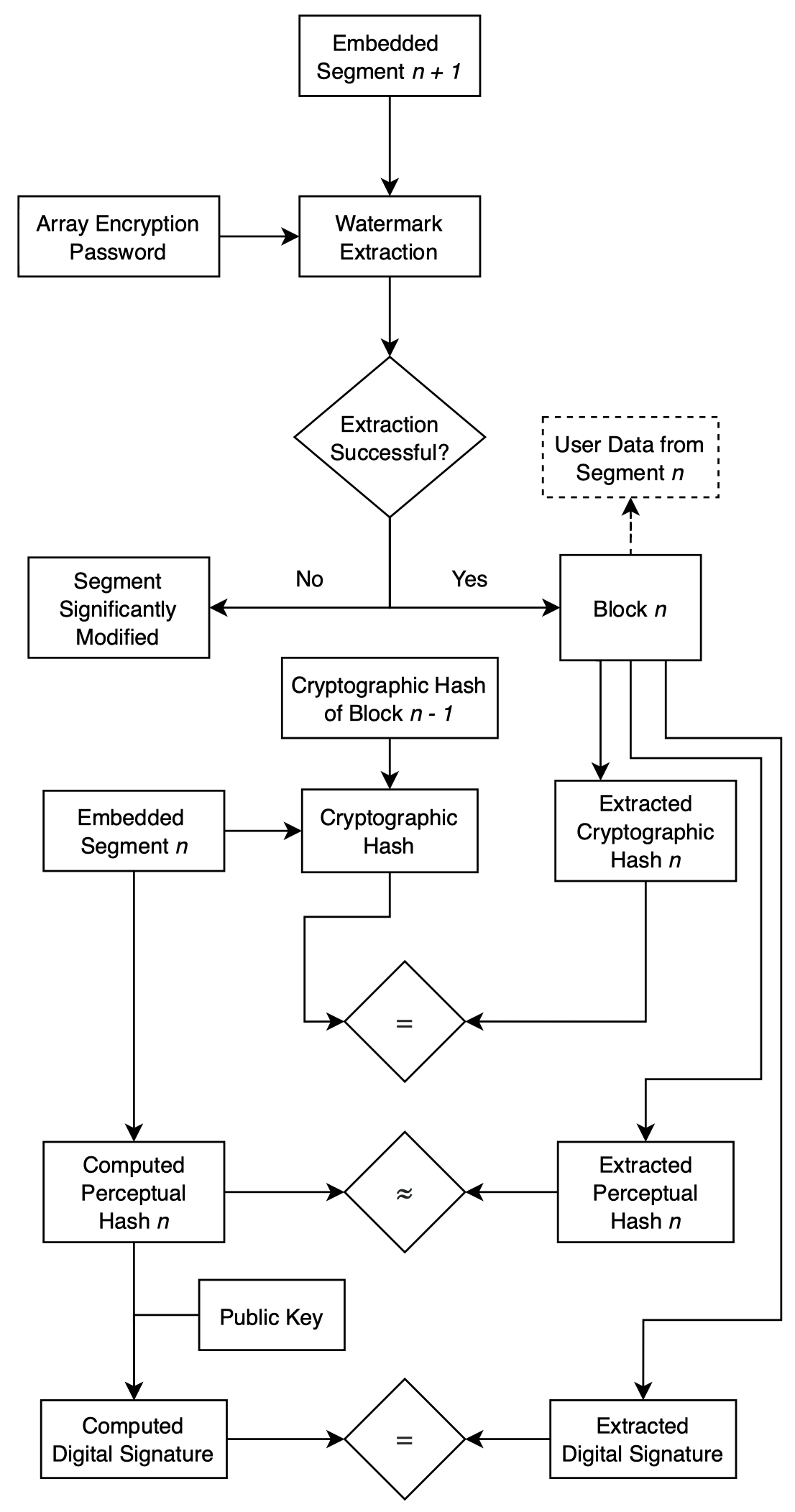}
\caption{A conceptual diagram of verifying one block.}
\end{figure}

When extracting the block from the watermark, the same watermarking parameters used in the embedding 
process must be used. \\

There is no requirement that the user data is static from block to block, and may be accumulated 
block--by--block and bit--by--bit to store larger amounts of data.

\subsection{Practical Considerations}

We will now detail some practical considerations for the design and implementation of an Embedded 
Blockchain. These include
\begin{enumerate}
	\item \textbf{The hash size}. The main consideration with the hash size is the need to balance the security of the 
	cryptographic hash and digital signature, the fidelity of the perceptual hash, and the size of the 
	user data embedded in each block with the size of the hidden message embedded in the watermark. 
	\item \textbf{The segment size}. Increasing the segment size decreases the temporal resolution of each 
	block, however, it increases the space to store the block. Increasing the segment size is useful 
	when embedding in a format which aggressively compresses the media. Increasing the segment size also 
	decreases the computational work required to create the Embedded Blockchain.
	\item \textbf{Comparing perceptual hashes}. Absolute numerical equality would be 
	desirable, however when using lossy compression to store the media often a threshold needs to be 
	estimated (as a function of the media size, the segment size and the bitrate of the perceptual hash). 
	\item \textbf{The size and dimensionality of the watermarking arrays}. Increasing the size of the arrays 
	also increases the SNR of the block extraction. Increasing the dimensionality also increases the 
	size of the hidden message. On the other hand, increasing the dimensionality also decreases the bit-rate 
	of each shift. 
	\item \textbf{The embedding domain of the watermark}. Common choices are the spatial and Fourier domains, 
	however, there are many possible choices \cite{cheng2003}\cite{gunjal2010}\cite{tao2004}. The choice of embedding 
	domain will affect the perceptibility and SNR of the watermark extraction. 
	\item \textbf{The user data}. Increasing the size of user data increases the 
	size of the block and subsequently the number of watermarking arrays required to embed, which decreases 
	the SNR of the watermark extraction.  
\end{enumerate}

\subsection{An Exemplar Calculation of the Block Size for Compressed Video}

A key software engineering challenge of the Embedded Blockchain is the data--carrying capacity of 
the watermark in the presence of lossy video compression schemes, whilst remaining visually 
imperceptible. \\

We will now detail a calculation which estimates the data capacity of the 
watermark for a single segment of 4K resolution video (3840 by 2160 px). We have embedded a 
family of 2D arrays of size $1087 \times 1087$ in a $2 \times 4$ tiled arrangement. These arrays 
possess good autocorrelation and pairwise cross--correlation properties \cite{blake2020}. The 
watermark embedding was in the spatial domain and the segment size (the number of frames described 
in a single block) was 15, or approximately half--second blocks. \\

Each 2D watermarking array can hold 2 integers in base 1087. If we allocate 3, 
256--bit hashes for the digital signature, perceptual hash, cryptographic hash and 256--bits for 
user data in each block, then we must store 1024 bits in the watermark (per block). We can 
bit--pack the 1024 bits into 104 base 1087 integers, or 52, 2D arrays. \\

In the following graphic, we have plotted the mean SNR of the watermark extraction with the 
size of the hidden message. 

\begin{figure}[H]
\label{wm_snr_v_strength}
\centering
\includegraphics[width=0.45\textwidth]{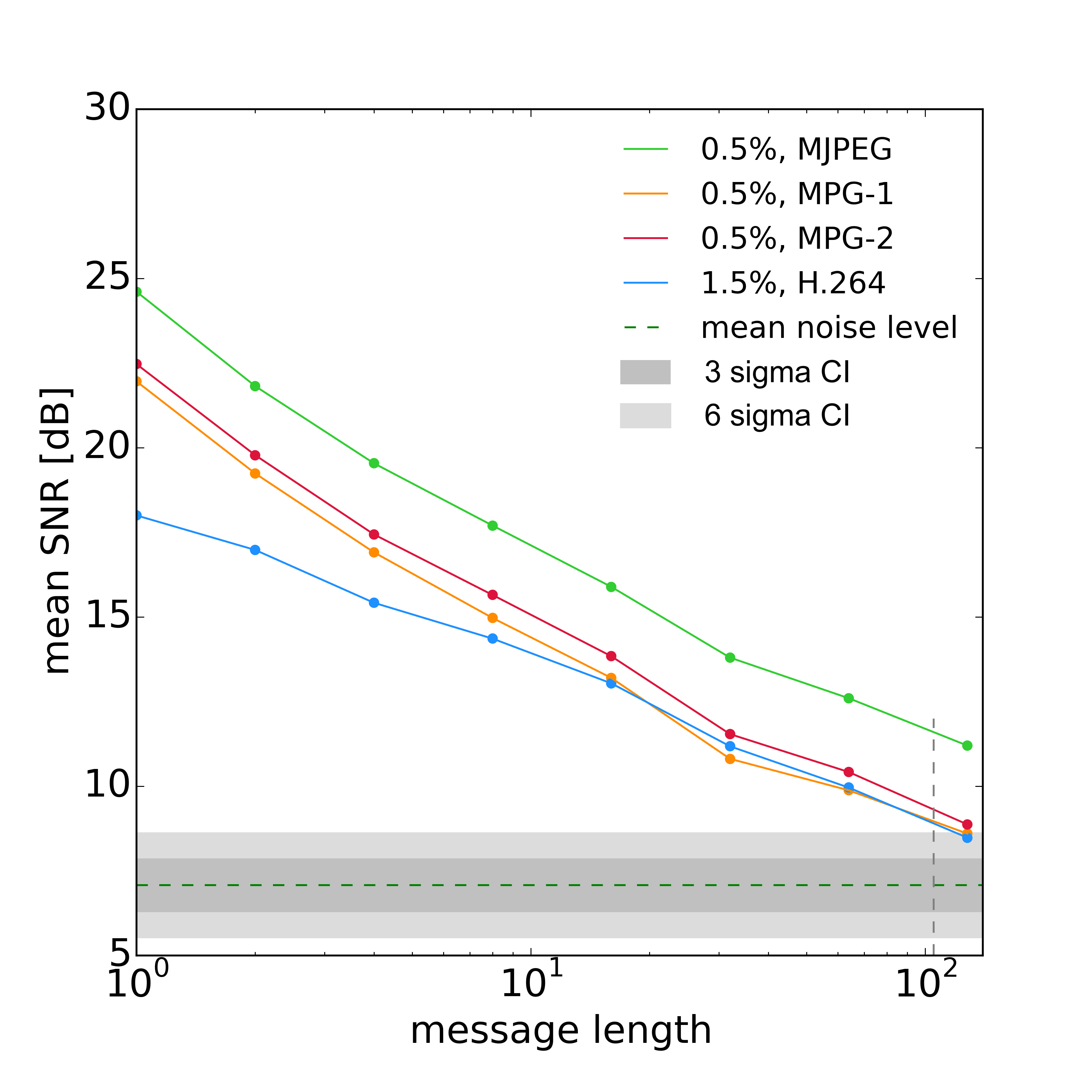}
\caption{A comparison of the watermark extraction SNR with the length of the hidden message. The 
compression schemes used were MJPEG, MPEG--1, MPEG--2 and H.264. The estimated SNR 
of random noise and the 3-- and 6--sigma confidence intervals are shown.}
\end{figure}

This calculation indicates a 1024 bit block size is feasible, as the SNR for a message length 
of 104 is above the 6--sigma confidence interval from the mean noise level. For MJPEG, 
MPEG--1 and MPEG--2 we achieved this message length with an embedding strength of 0.5\%, 
whereas for H.264 we needed to increase the embedding strength to 1.5\%. The required increase in 
embedding strength is due to the aggressive compression used in the H.264 scheme.

\section{Conclusions} 

We have described a two--step scheme to detect manipulations in media including video and audio. If 
efficiently implemented, this scheme should be able to run passively on a battery--powered device 
like a smartphone when capturing a video or audio recording. If so, it is likely that in the future all 
recording devices will contain an Embedded Blockchain--like scheme for proof of ownership and 
content verification. \\


\section{Acknowledgements}
I would like to thank Andrew Tirkel for many enlightening conversations on digital 
watermarking and The University of Melbourne for access to the Spartan high 
performance computing system \cite{lafayette2016}. 

\small
\bibliographystyle{abbrv}

\end{multicols}

\end{document}